\documentclass[5p]{elsarticle}

\usepackage{amssymb}

\usepackage{graphicx}
\usepackage{amssymb}
\journal{Physica B}

\begin{document}

\begin{frontmatter}

 \title{Na order and Co charge disproportionation in Na$_{x}$CoO$_{2}$}
 
 \author[label1,label2]{I.R.~Mukhamedshin}
 \ead{Irek.Mukhamedshin@kpfu.ru} 
 \author[label2]{H.~Alloul}
 \ead{alloul@lps.u-psud.fr} 
 \address[label1]{Institute of Physics, Kazan Federal University, 420008 Kazan, Russia}
 \address[label2]{Laboratoire de Physique des Solides, CNRS UMR 8502, Universit\'e Paris-Sud, 91405 Orsay, France, EU}

\begin{abstract}
We have synthesized and characterized different stable phases of sodium cobaltates Na$_{x}$CoO$_{2}$ with sodium content $0.65<x<0.80$. We demonstrate that $^{23}$Na NMR allows to determine the difference in the susceptibility of the phases and reveals the presence of Na order in each phase. $^{59}$Co NMR experiments give clear evidence that Co charge disproportionation is a dominant feature of Na cobaltates. Only a small fraction ($\approx$ 25\%) of cobalts are in a non-magnetic Co$^{3+}$ charge state whereas electrons delocalize on the other cobalts. The magnetic and charge properties of the different Co sites are highly correlated with each other as their magnetic shift $K_{ZZ}$ scales linearly with their  quadrupolar frequency $nu_Q$. This reflects the fact that the hole content on the Co orbitals varies from site to site. The unusual charge differentiation found in this system calls for better theoretical understanding  of the incidence of the Na atomic order on the electronic structures of these compounds. 
\end{abstract}

\end{frontmatter}

\section{Introduction}
\label{Introduction}
The influence of the dopant atoms on the electronic properties of conducting layers has initiated large debates in High Temperature Superconducting (HTSC) cuprates as well as in other complex layered oxides of transition elements. Whereas in many systems this influence is masked by miscellaneous effects, there are many experimental evidences that in the sodium cobaltates Na$_{x}$CoO$_{2}$ a large interplay between the Na atomic ordering and the electronic density on the Co sites occurs. 

As HTSC cuprates the sodium cobaltates Na$_{x}$CoO$_{2}$ are layered oxide materials and the charge doping of the CoO$_{2}$ layers is controlled on a large range by variation of the Na content. This can be put in parallel with the doping of the cuprates by chemical substitutions on the layers separating the CuO$_{2}$ planes. One significant difference with the cuprates is that the Co of the CoO$_{2}$ plane are ordered on a triangular lattice and not on a square lattice as for the CuO$_{2}$ plane of the cuprates. In this configuration the large crystal field on the Co site favors a low spin state \cite{SinghPRB61} in which orbital degeneracy influences significantly the electronic properties and may yield large thermoelectric effects \cite{TerasakiTEP}. A rich variety of other physical properties ranging from ordered magnetic states \cite{Mendels05}, high Curie-Weiss magnetism and metal insulator transition \cite{Foo}, superconductivity \cite{TakadaNature} \emph{etc} have then been observed on the cobaltates. These differences with the cuprates have apparently recently stimulated an increased interest of theorists as monitored by recent publications \cite{Karim2014,Chubukov2014,Wilhelm2014}.

In sodium cobaltates the low spin configuration for the Co ions should corresponds to 3$^{+}$ or 4$^{+}$ charged states with spin $S=0$ and $S=\frac{1}{2}$ respectively. Many experiments and theoretical calculations have considered that the Co magnetism either is uniform or that there is an Co$^{3+}$/Co$^{4+}$ charge segregation with localized magnetic moments. However it has been evidenced by NMR by many groups that for $x\geq 1/2$ the Na$^{+}$ displays an atomic ordering associated with Co charge disproportionation and itinerant magnetism in the planes  \cite{CoPaper,ImaiPRL1,Ray,Gavilano1,Ishida07,EPL2008,MHJulien075,H67_CoNMR}. Na ordered atomic structures have been observed in sodium cobaltates by TEM \cite{Zandbergen}, neutrons \cite{Roger}, and x-rays \cite{TaiwanPRB2009,H67NQRprb,FouryPRB}. However, the experimental situation that prevails so far is quite unusual in solid state physics, as most experiments do not permit altogether to establish reliably the relation between the local order proposed, the actual Na content and the local magnetic properties of the studied samples.  

On the contrary $^{23}$Na and $^{59}$Co NMR experiments have proved to be excellent probes allowing to evidence not only the Na atomic order but also that Co charge disproportionation occurs. In this paper we summarize our NMR study of sodium cobaltates Na$_{x}$CoO$_{2}$ with sodium content $0.65<x<0.80$ which allowed us to establish this correlation for the specific $x=2/3$ \cite{EPL2009,H67NQRprb,H67_CoNMR} and $x=0.77$ phases \cite{Na077prb,Co77JETP,Co077prb}. We also compare the properties of the studied phases with properties of the $x = 1$ and $x = 1/2$ compounds.

\section{Samples}
\label{Samples}

\begin{figure}[tbp]
\centering
\includegraphics[width=0.9\linewidth]{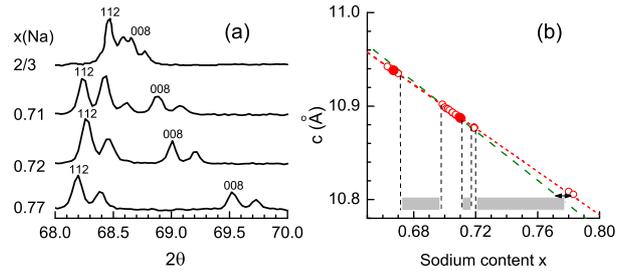}
\caption{(Color online) (a) Part of the x-ray powder spectra between 68 and 70$^{\circ }$ reflection angles for pure phase samples in the sodium content range $0.65<x<0.8$. One can see that in these pure phases the simple observation of the (008) Bragg peaks permits ensuring the absence of phase mixing. (The double peak structure corresponds here to the two Bragg peaks associated with the Cu K$\protect\alpha _{1}$ and K$\protect\alpha _{2}$ radiations); (b) The $c$ parameter values (filled and empty circles) obtained for all pure phase samples we could synthesize in the sodium content range $0.65<x<0.8$ establishes a  calibration curve for the $c$ axis parameter versus Na content for the $P2$ hexagonal two layers phases (red dotted line). Intervals with no data point marked by grey bars correspond to composition gaps. The $c(x)$ curve reported in Ref.~\cite{NatureMat2010} (green dashed line) agrees well with our data.}
\label{FigGrXrayEcrys}
\end{figure}

We have synthesized our powder samples by standard solid state reaction of Na$_{2}$CO$_{3}$ and Co$_{3}$O$_{4}$ powders in flowing oxygen, with nominal
concentrations $x$ increasing by increments of 2 to 3/1000 in the range
$0.65<x<0.80$. X-ray powder diffraction data always exhibited the Bragg peaks corresponding to the two layer Co structure $P2$ ($P6_{3}/mmc$, n$^{\circ }$166) with a hexagonal unit cell. We found that $c$-axis parameter changes systematically with sodium content $x$, so the position of the (008) reflection for this structure allows to match them (Fig.~\ref{FigGrXrayEcrys}). 

Also we systematically detected weak additional reflections indicative of three dimensional Na long range ordering. It has been immediately clear that the corresponding Na order is complicated and highly dependent of Na content. The systematic study of the powder x-ray spectra for various nominal Na contents allowed us to separate pure phases from multiphase samples with distinct $c$ axis parameters - the simple observation of the (008) Bragg peaks like in Fig.~\ref{FigGrXrayEcrys}(a) permits ensuring the absence of phase mixing. The concentrations for which single phase samples could be stabilized are sequenced in four distinct narrow $x$ non overlapping domains - see Fig.~\ref{FigGrXrayEcrys}(b). For intermediate concentrations between these, we always find mixed-phase samples with the two end phases. The crystallographic parameters for these 4 phases were reported in Ref.~\cite{EPL2008}. 

We have found that keeping samples at room temperature in contact with humid air leads to destruction of the phase purity and loss of sodium content \cite{JETP4phasesNQR}. The high sodium content sample evolves progressively into a mixture of the detected stable phases until it reaches the $x = 2/3$ composition which appears to be the most stable phase in this part of phase diagram.

Single crystals of sodium cobaltates Na$_{x}$CoO$_{2}$ can be grown by the floating zone technique \cite{SCgrowPRB2004,SCgrowJCG2004}. To perform detailed NMR study of the $x=0.77$ phase of sodium cobaltates we also synthesized single phase crystals with this specific sodium content \cite{Co77JETP}.

\section{Magnetic susceptibility and $^{23}$Na NMR shifts}

\begin{figure}[tbp]
\centering
\includegraphics[width=0.8\linewidth]{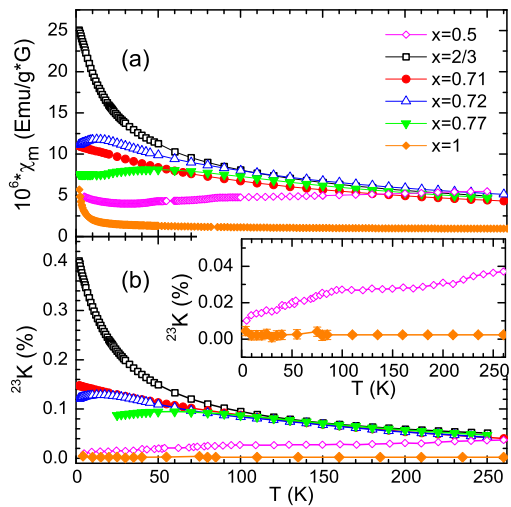}
\caption{(Color online) (a) $T$-dependencies of the bulk susceptibility $\chi_{m}$ measured in non-oriented powders of sodium cobaltates with $0.65<x<0.8$ with a DC SQUID in a 5T field; (b) $T$-dependence of the mean $^{23}$Na NMR magnetic shift. Identical behaviour above 100~K can be seen for the four phases with remarkable differences at low $T$. For comparison data for $x=0.5$ and $x=1$ phases also shown in panels (a) and (b), as well as in inset (data were taken from the Ref.~\cite{Bobroff05} and Ref.~\cite{LangNa1} correspondingly).}
\label{FigGrSuscEcrys}
\end{figure}

SQUID measurements of the macroscopic susceptibility $\chi _{m}$ taken in 5~T field allow us to evidence that the different phases display different magnetic properties. For instance, as evidenced in Fig.~\ref{FigGrSuscEcrys}(a), the low T magnitude of $\chi_{m}$ decreases progressively with increasing $x$. The $x=0.77$ phase is furthermore found to be the only phase in which an antiferromagnetic (AF) order is detected in low applied field. However, as minute amounts of impurities or slight admixture of phases could spoil the bulk measurements, spectroscopic measurements with local probes give better determination of the susceptibility of each phase.

The magnetic properties of the compounds are probed at the local scale through the magnetic shifts of the different Na sites resolved in the central line ($-\frac{1}{2}\leftrightarrow \frac{1}{2}$ transition) of the $^{23}$Na NMR spectra presented in Fig.~\ref{FigGrNaSpecs}. As we have shown in Ref.~\cite{NaPaper} shift values $K$ which corresponds to the first moment (center of gravity) of the central line allows us to follow the spin susceptibility $\chi _{s}(T)$ of the system up to room $T$, even when the central lines of the different Na sites merge. The $T$ variations of $K$ are reported in Fig.~\ref{FigGrSuscEcrys}b, and are shown to be quite identical for $T>100$~K with a unique Curie-Weiss $(T-\Theta)^{-1}$ variation (with $\Theta\approx$-80~K). They differ markedly below 100~K, as does the SQUID data for $\chi _{m}$, the low $T$ enhancement of $\chi _{s}(T)$ observed for $x=2/3$ being progressively reduced for increasing $x$.

In the $x=0.77$ AF phase, the saturation of $K(T)$, that is $\chi _{s}^{\alpha }(T)$, seen at low $T$ in Fig.~\ref{FigGrSuscEcrys}b should be associated with the onset of AF correlations. In a uniform Heisenberg model, one would then assign the progressive increase of $\chi _{s}(T)$ at low $T$ with decreasing $x$ to a decrease of $T_{N}$ and of out of plane AF coupling strength. However this primary interpretation fails as NMR data taken down to 1.4~K (and $\mu $SR to 50~mK \cite{Mendels05}), did not evidence any frozen magnetic state in the three other phases, which are then paramagnets in their ground state, most probably metallic, as no electronic magnetic transition is detected.

For comparison we also show in Fig.~\ref{FigGrSuscEcrys} the corresponding data for $x=0.5$ and $x=1$ phases taken from the Ref.~\cite{Bobroff05} and Ref.~\cite{LangNa1} correspondingly. The bulk magnetic susceptibility of Na$_{1}$CoO$_{2}$ exhibits a small and flat susceptibility, with a significant increase only at low temperature which is attributed to spurious paramagnetic local moments - see Fig.~\ref{FigGrSuscEcrys}(a). The magnetic shift $^{23}$K value in this phase is extremely low with no measurable temperature dependence (Fig.~\ref{FigGrSuscEcrys}(b) and insert in it). Both these dependencies confirm that the Na$_{1}$CoO$_{2}$ composition corresponds to a band insulator made of filled shell nonmagnetic Co$^{3+}$.

In Fig.~\ref{FigGrSuscEcrys} the temperature dependencies of the bulk magnetic susceptibility and magnetic shift of Na$_{0.5}$CoO$_{2}$ compound are also displayed \cite{Bobroff05}. Both are very weakly temperature dependent as compared with  phases with sodium content $0.65<x<0.80$. Two kinks at $T$=51~K and $T$=86~K (see insert in Fig.~\ref{FigGrSuscEcrys}(b)) can be attributed to a metal-insulator transition (MIT) and long range antiferromagnetic order, correspondingly \cite{Foo}.

\section{$^{23}$Na NMR and Na order}

To perform the nuclear magnetic resonance (NMR) study of sodium cobaltates powders we prepared oriented samples. As the room $T$ susceptibility of sodium cobaltates is known to be anisotropic \cite{Wang}, we have used this anisotropy to align the single crystallites of powder samples in a 7~T field, by mixing the samples with Stycast 1266 epoxy resin which cured in the field. Thus in our samples the $c$  axes of the crystallites were aligned in the same direction, but the $ab$ planes of different crystallites are randomly distributed. Such Stycast epoxy matrix also perfectly protects the powder from environment (water) influence \cite{JETP4phasesNQR}. Samples with $x=2/3$ prepared almost nine years ago and packed in Stycast did not show any changes in the NMR spectra.

\begin{figure}[tbp]
\centering
\includegraphics[width=0.8\linewidth]{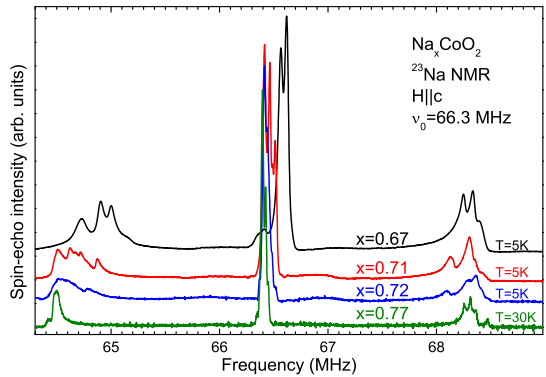}
\caption{(Color online) Low temperature $^{23}$Na NMR spectra for for pure phase oriented powder samples in the sodium content range $0.65<x<0.8$. The splitting of the satellite $^{23}$Na transitions reveals the existence of defined Na order at each specific sodium content. These NMR spectra were obtained by pulse NMR at fixed frequency $\nu _{0}$, the spin echo intensity being measured after two 2$~\mu$ sec radio frequency pulses separated by 100~$\mu$ sec, while the external magnetic field was increased by steps. The full NMR spectra were then constructed using a Fourier mapping algorithm \cite{Clark}.}
\label{FigGrNaSpecs}
\end{figure}

In Fig.~\ref{FigGrNaSpecs} the typical low $T$ $^{23}$Na NMR spectra for the pure phase oriented samples are shown. As $^{23}$Na has a spin $I=3/2$, the NMR spectrum for a single Na site displays a central transition and two satellites disposed symmetrically with respect to the central line. While the shift of the central line signal is governed by the magnetism of the near-neighbor Co sites, the distance $\Delta \nu $ between the satellites is linked to the quadrupole frequency $\nu _{Q}$ associated with the magnitude of the electric field gradient (EFG) at the Na site, which is governed by the distribution of ionic and electronic charges around the Na. For each Na site the largest splitting between satellites is observed when $H\parallel c$, which evidences that the principal axis $Z$ of the EFG tensor is close to the crystallographic $c$ axis for all sodiums. For $x=2/3$ phase in the spectra of Fig.~\ref{FigGrNaSpecs} one can clearly distinguish three pairs of quadrupole transitions which correspond to three Na sites with distinct local environments \cite{NaPaper}. Due to the more complex structures of the other phases there was no surprise in finding a larger number of resolved Na sites, although with similar magnitudes of their quadrupole splittings. For example detailed analysis done in the Ref.~\cite{Na077prb} for $x=0.77$ phase shows that there are 4 unequivalent Na position in this phase of sodium cobaltates. The fact that only limited number of sites are observed in each phase definitely proves that the neighboring charges of the Na sites are ordered.

\section{2D Na order}

\begin{figure}[tbp]
\centering
\includegraphics[width=0.7\linewidth]{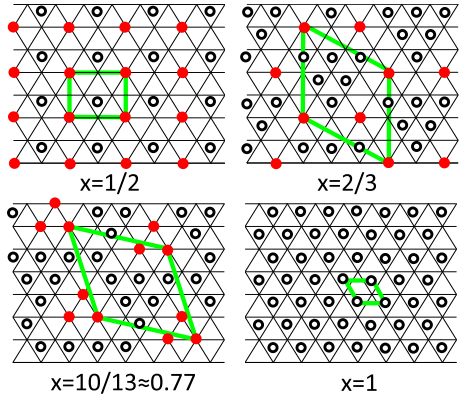}
\caption{(Color online) 2D Na unit cell for specific phases of sodium cobalates Na$_{x}$CoO$_{2}$ (thick green lines). The Na1 sites (filled red dots) and Na2 sites (empty black circles) are differentiated, and represented above the triangular lattice of Co sites (not reported) which are located at all intersections of thin black lines.} 
\label{FigStructures}
\end{figure}

Na$_{1}$CoO$_{2}$ has been previously synthesized and its structure determined on both powder samples \cite{Na1Fouassier} and single
crystals \cite{Na1Takahashi}. In this phase the single Na site is located on top of the center of a Co triangle, in a configuration usually called Na2 - see Fig.~\ref{FigStructures}. 

Numerical simulations \cite{Roger} and electronic structure calculations \cite{Hinuma} suggest that the Na2 vacancies formed for $x<1$, are ordered in the Na plane. These vacancies have a tendency toward clustering and, depending on $x$, these clusters induce altogether the appearance of isolated Na1 sites (on top of a Co) in divacancy clusters or of trimers of Na1 sites in trivacancy clusters \cite{Roger}. For $x=0.5$ the Na atoms are ordered in an orthorhombic superstructure commensurate with the Co lattice with the equal occupancy
of Na1 and Na2 sites \cite{Na05structure} (Fig.~\ref{FigStructures}). 

For the $x=2/3$ phase our NMR/NQR experiments have allowed the determination of both the 2D Na ordered structure and the 3D stacking of Na/Co planes \cite{EPL2009} (Fig.~\ref{FigStructures}). This Na organization agrees with GGA calculations \cite{Hinuma}, and consists of two Na on top of Co sites (the Na1 sites) and six on top of Co triangles (the Na2 sites). This twelve Co unit cell and the stacking between planes has been confirmed later by x-ray diffraction experiments \cite{H67NQRprb}. It results in a differentiation of four cobalt sites in the structure, two nearly non-magnetic Co$^{3+}$ and two more magnetic sites constituting a kagom\'e sublattice of the triangular Co lattice on which the holes are delocalized \cite{H67_CoNMR}.

The existence of a sodium site with axial charge symmetry and the intensity ratio between the sets of $^{23}$Na NMR lines in the sodium cobaltates phase with $T_{N}=$22~K permits us to reveal that the two-dimensional structure of the Na order corresponds to 10 Na sites on top of a 13 Co sites unit cell, that is with $x=10/13\approx 0.77$ \cite{Na077prb} - see Fig.~\ref{FigStructures}. However contrary to the $x=1/2$ or $x=2/3$ phases the 3D stacking of the Na planes is not perfect for $x=0.77$ \cite{Co077prb} but this does not influence markedly the electronic properties of this phase. 

\section{Co charge disproportionation}

NMR/NQR is a powerful technique which allows to establish the relation between the local Na order and the local magnetic properties of the phases studied. 

In the $x=1$ phase the $^{59}$Co NMR signal shows a single cobalt site with very slow nuclear relaxation \cite{LangNa1,MHJulienNa1}, as expected for homogeneous planes of Co$^{3+}$ ions with $S=0$. 

$^{59}$Co NMR in the $x=1/2$ phase gives clear evidence for the presence of two Co sites. Both sites display a similar $T$ dependence of the spin contribution to the magnetic shift, suggesting that there is no charge segregation into Co$^{3+}$ and Co$^{4+}$ sites. The electric field gradient at the Co site does not change at the AF or MIT transitions, indicating the absence of any modification of the charge ordering. The analysis of hyperfine coupling constants for both Co sites allowed the authors of \cite{Bobroff05} to conclude, that in this $x=1/2$ phase the charge disproportionation on the Co sites is rather small: Co$^{3.5\pm \varepsilon }$ (with $\varepsilon <0.2)$. 

\begin{figure}[tbp]
\centering
\includegraphics[width=0.8\linewidth]{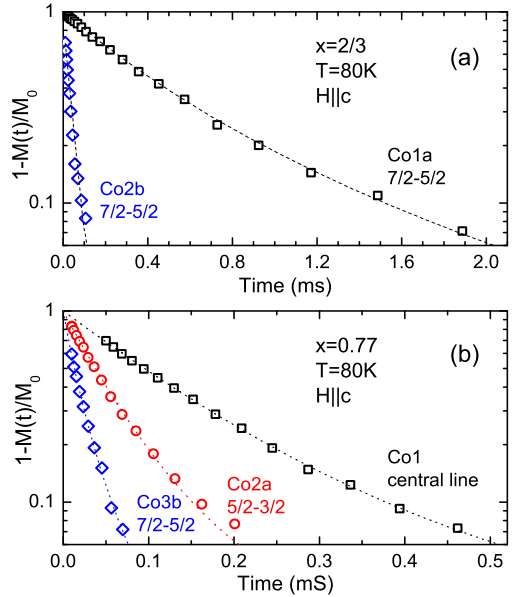}
\caption{(Color online) Spin-lattice relaxation curves measured on different
cobalt sites in the $x=2/3$ and $x=0.77$ pure phase samples in the same experimental conditions. Whereas in the $x=2/3$ there are two types of cobalts - slow (Co1a,b) and fast (Co2a,b) relaxing, in the $x=0.77$ phase there are three types - slow (Co1), intermediate (Co2a,b,c) and fast (Co3a,b) relaxing. The fits of the curves with the magnetization relaxation functions from Ref.~\cite{H67_CoNMR} with $T_1$ parameters from the Table~\ref{tab:CoParams} are also shown.}
\label{FigGrT1curves}
\end{figure}

Though $^{59}$Co NMR has been detected in many phases, the most well studied one has been the $x=2/3$ phase. As we have formerly established the in plane Na ordering and its three dimensional stacking in this phase, we succeed for the first time to completely resolve all the parameters of the Zeeman and quadrupolar Hamiltonians for all cobalt sites in the unit cell \cite{H67_CoNMR}. The three non-magnetic Co$^{3+}$ (Co1) in $x=2/3$ unit cell are in axially symmetric positions and the doped holes are delocalized on the nine complementary magnetic cobalt sites (Co2) of the atomic unit cell. This Co1/Co2 charge disproportionation can be clearly seen in the relaxation of nuclear magnetization which is slow for non-magnetic Co1 sites and fast for the Co2 magnetic sites. This is exemplified in Fig.~\ref{FigGrT1curves}(a) where the relaxation curves of the longitudinal nuclear magnetization (spin lattice relaxation) are shown for Co1a and Co2b sites.

However the moderately complicated atomic structure in the $x=2/3$ phase resumes then in a very simple electronic structure in which the electrons delocalize on the Co2 kagom\'e sublattice of the triangular lattice of Co sites. These sites display a strong in plane electronic anisotropy initially unexpected but which accords perfectly with an orbital ordering along the kagom\'e sublattice organization and could result from the distribution of holes between the axial $a_{1g}$ orbitals and the in plane $e_{g}^{\prime}$ orbitals on the Co2 sites \cite{H67_CoNMR}. The observation of a single temperature dependence of the spin susceptibilities indicates that a single band picture applies, and that the magnetic properties are dominated by the static and dynamic electronic properties at the Co2 sites.  

The $^{59}$Co NMR study in the paramagnetic state of the $x=0.77$ phase with $T_{N}=22$~K permitted us to evidence that at least 6 Co sites could be differentiated \cite{Co77JETP}. They could be separated by their magnetic and relaxation behavior into three types: a single site (Co1) with cobalt close to non-magnetic Co$^{3+}$, two sites (Co3) with the most magnetic cobalts in the system, and the remaining three sites (Co2) displaying an intermediate behavior - see Fig.~\ref{FigGrT1curves}(b). 

It is important to highlight that the non-magnetic Co1 sites correspond only to 25\% of total cobalts in the system for $x=2/3$ and 23\% for $x=0.77$. These facts are in complete contradiction with Co$^{3+}$/Co$^{4+}$ charge segregation model, for which the Co$^{3+}$ content should be proportional to the Na content.

\section{Estimation of the charge disproportionation for the different Co
sites}

\begin{figure}[tbp]
\centering
\includegraphics[width=0.8\linewidth]{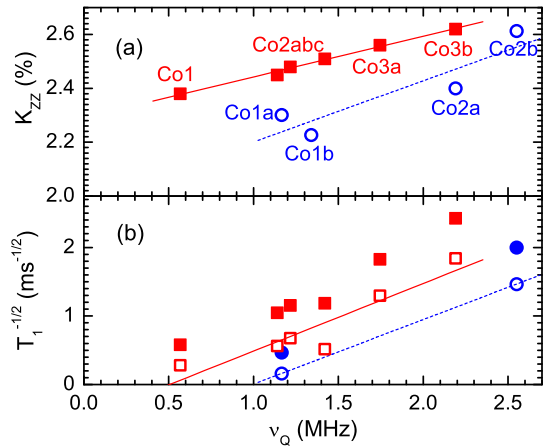}
\caption{(Color online) (a) Magnetic shift $K_{ZZ}$, (b) $(1/T_{1})^{1/2}$ (closed symbols) and the corresponding pure spin contribution  $(1/T_{1spin})^{1/2}$ (open symbols) vs. $\protect \nu _{Q}$ for cobalt sites in the phases with $x=0.77$ (red squares) and $x=2/3$ (blue circles). All experimental points were obtained at T=80~K. Linear fits are shown as guides to the eyes.}
\label{FigCoCorrEcrys}
\end{figure}

Our $^{59}$Co NMR experiments in the $x=2/3$ and $x=0.77$ phases allowed us to determine the quadrupole frequencies $\nu_{Q}$, magnetic shifts $K_{ZZ}$ and nuclear spin lattice relaxation rates $1/T_{1}$ for distinct cobalt sites in each phase \cite{H67_CoNMR,Co077prb}. They are summarized in Table~\ref{tab:CoParams}. Unexpectedly we found that the magnetic and charge properties of the Co sites are highly correlated with each other as $K_{ZZ}$ and $(1/T_{1})^{1/2}$ scale linearly with $\nu _{Q}$ - see Fig.~\ref{FigCoCorrEcrys}. 

\begin{table}[tbp]
\centering
\caption{Some NMR parameters of the different Co sites in the $x=2/3$ and  $x=0.77$ phases measured at $T$=80~K. $K_{ZZ}$ is the magnetic shift, $\protect\nu _{Q}$ is the quadrupolar frequency, $I_{n}$ is the number of cobalts corresponding to the given Co site, $T_{1}$ is the values of the longitudinal nuclear magnetisation relaxation time, $\protect\delta $ is the hole concentration (charge state of cobalt is Co$^{3+\protect\delta }$) deduced from the analysis of the correlation between $(1/T_{1 spin})^{1/2}$ and $\protect\nu _{Q}$ data.}
\label{tab:CoParams}%
\begin{tabular}{|c|c|c|c|c|c|}
\hline
     & $K_{ZZ}$, \% & $\nu_Q$, MHz & $I_{n}$ & $T_1$, ms & $\delta $ \\
\hline
\multicolumn{6}{|c|}{$x$=2/3} \\ 
\hline
Co1a & 2.30(1) & 1.165(5)& 2 & 4.59(2) & 0.05 \\
Co1b & 2.23(1) & 1.34(1) & 1 &         & 0.10 \\
Co2a & 2.40(1) & 2.19(1) & 3 &         & 0.35 \\
Co2b & 2.61(1) & 2.55(1) & 6 & 0.25(1) & 0.46 \\
\hline
\multicolumn{6}{|c|}{$x$=0.77} \\ 
\hline
Co1  & 2.38(1) & 0.568(5) & 3 & 2.98(3) & 0.02 \\
Co2a & 2.45(1) & 1.14(2)  & 4 & 0.91(5) & 0.20 \\
Co2b & 2.48(1) & 1.22(2)  & 1 & 0.75(4) & 0.23 \\
Co2c & 2.50(1) & 1.42(2)  & 2 & 0.71(4) & 0.29 \\
Co3a & 2.56(1) & 1.75(2)  & 2 & 0.30(1) & 0.39 \\
Co3b & 2.62(1) & 2.19(2)  & 1 & 0.17(1) & 0.53 \\
\hline
\end{tabular}
\end{table}

Such correlations can be understood if one takes into account that both measured values $\nu _{Q}$ and $K_{ZZ}$ contain terms proportional to the quadrupole moment of the $t_{2g}$ hole density distribution $\hat{q}$ which involves as a coefficient the hole concentration $\delta $ on the Co site \cite{Co077prb}. Therefore the linear variation with $\nu _{Q}$ found for the Knight shift $K_{ZZ}$ and $(T_{1})^{-1/2}$ just reflects the fact that the hole content on the Co orbitals varies from site to site in both phases. Also this observation means that the hyperfine coupling, or the local magnetic $\chi $ scales with the on site delocalized charge.

As observed experimentally for non-magnetic Co$^{3+}$ the spin lattice relaxation is extremely long \cite{LangNa1,MHJulienNa1} as the spin contribution $(1/T_{1spin})$ becomes negligibly small. Therefore in the plot of Fig.~\ref{FigCoCorrEcrys}(b) the pure Co$^{3+}$ state should correspond to the limit $(1/T_{1spin})^{1/2}$ $\rightarrow 0$. In that limit the corresponding $\nu _{Q}=\nu _{0}$ should be solely due to the lattice contribution $\nu _{Q}^{latt}$ to the EFG. We can assume that on each Co site the quantity $\nu _{Q}-\nu _{0}$ is directly related with the local charge contribution to the EFG, which is proportional to the hole concentration. Using the charge neutrality equation and knowing the number of cobalts corresponding to the given Co site we could estimate the hole concentration on each Co site in both phases - see last column in the Table~\ref{tab:CoParams} and Ref.~\cite{Co077prb} for details. 

This analysis permitted us to deduce a variation of $\nu _{Q}$ with local charge $d\nu_{Q}/d\delta \approx$3.3~MHz per hole on the Co site for both phases. 

We can apply our model to the in $x=1/2$ phase, in which a small charge disproportionation into Co$^{3.5\pm \varepsilon }$ is observed \cite{Bobroff05}. Using the measured $\nu _{Q}$ values 2.8~MHz and 4~MHz for the two Co sites in the $x=1/2$ phase we can estimate in our model that $\varepsilon =0.18$ which is in good agreement with the conclusion of Ref.~\cite{Bobroff05}.

The occurrence of a local charge contribution to the EFG of $d\nu_{Q}/d\delta $=3.3~MHz per hole on Co site implies that the local charge is not evenly distributed on the Co $t_{2g}$ orbitals. A theoretical estimate gives that single hole (or electron) on a given $t_{2g}$ orbital should correspond for $^{59}$Co to  $d\nu_{Q}/d\delta \approx$20~MHz per hole which is about six times larger than our experimental value \cite{Co077prb}. This allows us to suggest that in the cobaltates the holes are distributed over the three orbital states within the $t_{2g}$-shells due to the thermal/quantum fluctuations \cite{KHA2008}.

\section{Conclusion}

All NMR studies done so far have demonstrated that the layered Na cobaltates
undergo a charge disproportionation of the Co sites for large Na content. This has been ascertained by the observation of a slow relaxing $^{59}$Co NMR signal, that we call Co1, with an intensity which increased with hole content from $x=0.5$ to $0.8$. All experimental parameters, e.g.  magnetic shift or EFG allowed us to establish that those Co sites have nearly filled t$_{2g}$ orbitals that correspond to a Co$^{3+}$ with low spin (S=0) configuration. The NMR signals of the other Co sites have been much harder to study in detail as they usually display the metallicity with larger magnetic behavior which is detected in the macroscopic properties. 

In our recent studies we have highlighted two specific cases in which the $^{59}$Co NMR spectra could be resolved completely, which permitted us to achieve a complete study of the charge disproportionation. For $x=2/3$ which is a paramagnet with a Curie Weiss like susceptibility increasing down to $T=0$, only a second Co site (Co2) with a different magnetic behaviour is detected, while for $x=0.77$, which is AF below $T_{N}=$22~K, at least two other types of disproportionated Co sites are distinguished. In either case the observed charge disproportionation does not correspond to the  segregation into Co$^{3+}$ and Co$^{4+}$ often proposed by theoreticians. The physical properties do not correspond either to a situation where a bath of conduction electrons would be diffused by local moments responsible for the magnetic properties. We rather showed that the Co$^{3+}$ sites are inert magnetically, while the bands associated with the other sites correspond roughly to Co$^{3.5+}$ and are involved in both
metallic and magnetic properties. 

Let us point out altogether that these compounds display an ordered Na atomic structure for a limited number of $x$ values, and that phase mixing occurs for $x$ values in between two pure phases. In most of the pure phases the Na atomic order seems to be locked on that of the charge disproportionation of Co sites. However we cannot decide so far from the experimental side whether the Na order drives the charge disproportionation. The latter could as well be intrinsic to the electronic properties of the Co plane for a given charge filling and could conversely drive the pinning of the Na in an ordered structure.

Some theoretical studies favor the latter approach and indicated for instance that the kagom\'e structure of the Co2 sites found for $x=2/3$ could be intrinsic to the CoO$_{2}$ planes \cite{KoshibaeMaekawa}. The fact that we could not detect a 3D ordering of the Na in the $x=0.77$ phase would possibly support such a picture in which the disproportionation is specific to the Co layers. We certainly expect theoretical calculations involving electronic correlations to establish whether the  charge disproportionated state could occur in the absence of the Na potential. 

Such a study on cobaltates has some analogy as well with those performed on
the cuprates, where charge density wave order has been detected and heavily
studied recently in YBCO$_{6+x}$. The most prominent static CDW have been
detected there for O contents $x$ for which the dopants are ordered on the
CuO chains in between the CuO$_{2}$ bilayers. This suggests that in
layered oxides the ordering of the dopants has indeed an incidence on the
Fermi surface of the layers and therefore on their electronic properties.

\section{Acknowledgments}

We would like to thank here F.~Bert, P.~Mendels and A.V.~Dooglav for their
help on the experimental NMR techniques and for constant interest and
stimulating discussions. This work was partially supported by the RFBR under project 14-02-01213a and performed according to the Russian Government Program of Competitive Growth of Kazan Federal University. I.R.M. thanks for the support of a visit to Orsay by ``Investissements d'Avenir'' LabEx PALM (ANR-10-LABX-0039-PALM).



\bibliographystyle{elsarticle-num} 
\bibliography{NaxCoO2} 

\end{document}